\documentclass[prl,showpacs,twocolumn,nobalancelastpage,superscriptaddress,floatfix,amsmath]{revtex4}
\usepackage[latin1]{inputenc}
\usepackage[english]{babel}
\usepackage{graphicx,psfrag}
\usepackage{amsmath}
\usepackage{amssymb}
\usepackage{amscd}
\usepackage{eucal}
\usepackage{color}
\usepackage{bm}

\newcommand{\rmd}{{\rm d}}

\input{epsf}
\begin{document}
\title{Superconductivity-induced macroscopic resonant tunneling}
\author{M.C.~Goorden}
\affiliation{D\'epartement de Physique Th\'eorique,
Universit\'e de Gen\`eve, CH-1211 Gen\`eve 4, Switzerland}
\author{Ph.~Jacquod}
\affiliation{Physics Department,
   University of Arizona, 1118 E. 4$^{\rm th}$ Street, Tucson, AZ 85721, USA}
\author{J.~Weiss}
\affiliation{Physics Department,
   University of Arizona, 1118 E. 4$^{\rm th}$ Street, Tucson, AZ 85721, USA}
\date{\today}
\begin{abstract}
We show analytically and by numerical simulations that 
the conductance through $\pi$-biased chaotic Josephson junctions 
is enhanced by several orders of magnitude in the short-wavelength regime. 
We identify the mechanism behind this effect as
{\it macroscopic resonant tunneling} through a macroscopic number
of low-energy quasi-degenerate Andreev levels. 
\end{abstract}
\maketitle{}

Resonant tunneling is the process by which quantum tunneling
is enhanced by resonant transfer through intermediate 
quasi-bound states~\cite{Chang91}. The paradigmatic 
example is a one-dimensional double-barrier structure, where
the transmission probability is given by
\begin{equation}\label{eq:res_tunnel}
T_{\rm res}(E) = \frac{T_1 \, T_2}{1+R_1 \, R_2
-2 \sqrt{R_1 \, R_2} \cos[\Theta(E)]},
\end{equation}
in terms of the transmission and reflection 
probabilities $T_{1,2}=1-R_{1,2}$ of the
individual barriers. In the tunneling regime,
$T_{1,2} \ll 1$, narrow quasi-bound states
exist between the two barriers, with well resolved quantized 
energies, $\epsilon_m$. When the energy of the tunneling particle
coincides with one of these energies, 
$\Theta(E=\epsilon_m)=0$, and in the case of symmetric barriers,
$T_1=T_2$, the transmission is perfect, $T_{\rm res}(\epsilon_m)=1$.
This is to be contrasted with the transmission probability
$T(E) = T_1 T_2/4$ away from resonance,
and the incoherent transmission probability
$T(E) = T_1 T_2/(T_1+T_2)$ one obtains when inelastic scattering
occurs between the two barriers~\cite{But88}. 

Resonant tunneling also occurs in higher dimensions. In chaotic systems
with no spatial symmetry,
there is no degeneracy of the intermediate states. Therefore, considering
linear transport at low temperature, resonance occurs with at most one 
intermediate state at a time, leading at best to an increase of the average
conductance by an amount
$G_0 = 2 e^2/h$ -- it is a microscopic effect of order one.
In this article, we show that the proximity of the intermediate system
to two superconductors can lead to 
a totally different phenomenology, where 
resonant tunneling through a macroscopic number 
$\propto N_{\rm n}$ of intermediate levels occurs at the Fermi energy.
This results in a conductance $G \propto G_0 N_{\rm n}$ at resonance,
much larger than the nonresonant 
conductance $\propto G_0 \Gamma_{\rm n} N_{\rm n}$.
The resonance condition is met when the phase difference between
the two superconductors is $\phi=\pi$. We foresee that
this {\it macroscopic resonant tunneling} effect might have applications in 
current switching devices and magnetic flux ``transistors''.

\begin{figure}[ht]
\begin{center}
\psfrag{Gamma}{$\Gamma_{\rm n}, N_{\rm n}$}
\psfrag{gs1}{$\gamma_{s1}$}
\psfrag{gs2}{$\gamma_{s2}$}
\psfrag{gs3}{$\gamma_{s3}$}
\psfrag{gn}{$\gamma_{n}$}
\psfrag{R} {$R$}
\psfrag{L} {$L$}
\psfrag{S1} {$S_1$}
\psfrag{S2} {$S_2$}
\psfrag{D1}{$\Delta e^{-i \phi/2}$}
\psfrag{D2}{$\Delta e^{ i \phi/2}$}
\psfrag{Gammas}{$\Gamma_{\rm n}, N_{\rm s}$}
\includegraphics[width=9.5cm]{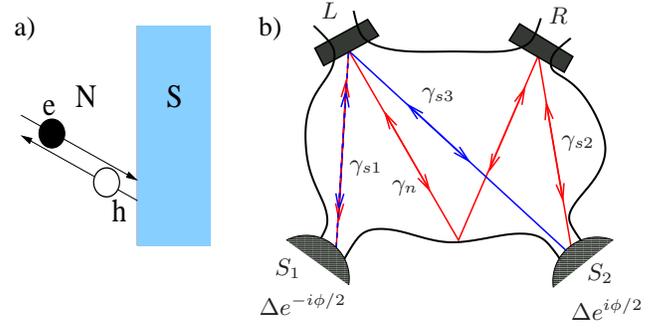}
\vspace{-1.1cm}
\end{center}
\caption{\label{fig:device}(Color online) a) Andreev reflection by a 
superconductor (S). An incoming electron (e) in a
normal metal (N) near the Fermi energy $E_{\rm F}$ is reflected as a hole (h) with
opposite velocity. b)
Schematic of our system. A ballistic 
metallic cavity is connected to two 
normal ($L$ and $R$) metallic leads, and 
two superconducting ($S_1$ and $S_2$) leads. The superconducting pair
potentials have a phase
difference of $\phi$. Two periodic Andreev 
orbits contributing
to macroscopic resonant tunneling at $\phi=\pi$ are drawn.\\[-5mm]}
\end{figure}

The system we investigate is sketched in Fig.~\ref{fig:device}.
A ballistic metallic quantum dot is connected to two
metallic electrodes
($L$ and $R$, each 
carrying $N_{\rm n} \gg 1$ channels) and two superconducting electrodes 
($S_1$ and $S_2$, each carrying $N_{\rm s} \gg 1$ channels). 
The electrodes are coupled to the
dot via tunnel contacts of transparency $0 < \Gamma_{\rm n,s} \le 1$, 
such that $1 \lesssim \Gamma_{\rm n} N_{\rm n} \ll N_{\rm n}$, and
$\Gamma_{\rm s} N_{\rm s} \gg 1$. 
We are interested in transport between the two normal leads as a function
of the phase difference $\phi$ 
between the two superconductor's pair potentials,
$\Delta_{S_1} = \Delta_{S_2} \exp[-i \phi]$, $|\Delta_{S_{1,2}}|=\Delta$.

The physics in our system is to a large extent governed by
Andreev reflection~\cite{And64}. At low energy, this is the dominant
reflection process at an interface between a metal
and a superconductor, where an electron is retroreflected
into a hole, and vice-versa. The process is sketched in 
Fig.~\ref{fig:device}. When the excitation energy $\epsilon$
is negligible against $\Delta$ 
and the Fermi energy $E_{\rm F}$, 
the retroreflection is perfect and the hole (electron)
exactly retraces the path previously followed by the electron (hole), with 
an additional Andreev reflection phase of $-\pi/2 \pm \phi/2$. 
In absence of normal lead (one then has 
an {\it Andreev billiard}~\cite{Kos95}), Andreev reflection renders all 
classical paths periodic in a cavity that would be chaotic otherwise.

When the cavity is weakly connected to 
external leads, transport can be resonantly mediated through those periodic
orbits that touch the contacts to the leads and both superconductors. 
Two such orbits are depicted
in Fig.~\ref{fig:device}. Each such orbit represents a family of
scattering trajectories 
constructed from a {\it primitive trajectory}, and an {\it Andreev loop}
that can be traveled $p$ times, $p=0,1,2,\ldots$. 
We indicate segments of trajectories as $\gamma_a^{(\alpha)}$ with a 
superscript
$\alpha=e, \, h$ denoting whether an electron or a hole travels on the
trajectory, and a subscript identifying whether the segment touches a superconducting
lead ($a=s_i$) or only normal leads ($a=n$).
With this convention, the blue trajectory in Fig.~\ref{fig:device} 
represents trajectories coded in Eq.~(\ref{eq:classI}). 
At $\epsilon=0$, the action phase 
accumulated along such trajectories is given by 
$
S_\gamma(p) = p (-\pi \pm \phi) + \varphi,
$
where $\varphi$ is a constant phase that is irrelevant for transport.
The crucial element is that the phase difference between the two 
superconductors cancels out the accumulated Andreev scattering phase
when $\pi=\phi$. Then all members of a family
interfere constructively with each other, because
$S_\gamma(p)-S_\gamma(p')=0$. 
This holds simultaneously for {\it all} families of 
trajectories that touch both superconductors,
with the topology of the
trajectories sketched in 
Fig.~\ref{fig:device}~\cite{caveat}. As there are infinitely many such 
trajectories, the result is macroscopic resonant tunneling with a conductance
$G(\phi=\pi) \propto G_0
N_{\rm n}$, independent of $\Gamma_{\rm n} \ll 1$. 
Neither macroscopic resonant tunneling, nor the associated
massive quasi-degeneracy of energy levels of  
Andreev billiards around $\epsilon=0$ for $\phi=\pi$
have
been noticed in earlier investigations of 
the density of states
of Andreev billiards~\cite{Mel97,Zho98,Lod98,Jac03,Kor04} and of
transport through the Josephson junction set-up of 
Fig.~\ref{fig:device}~\cite{Spi82,Bee95,Kad99}.

We sketch our analytical calculation.
In the symmetric configuration we consider, where
each normal lead carries the same number of channels, connected
to the cavity with the same transparency, the
average conductance from $L$ to $R$
reads, to leading order in $\Gamma_{\rm n} N_{\rm n}$~\cite{Lam93}
\begin{equation}\label{eq:sym_lambert}
\langle G \rangle /G_0 = \langle T_{RL}^{ee} \rangle + \langle T_{LL}^{he} 
\rangle.
\end{equation}
Here, $T_{ji}^{\beta \alpha}$ gives the transmission probability for a
quasi-particle of type $\alpha$ from the normal lead $i$ to a quasi-particle of
type $\beta$ into the normal lead $j$. 
To evaluate the resonant contributions to $\langle T_{RL}^{ee} \rangle $ 
and $\langle T_{LL}^{he} \rangle$,
we follow the semiclassical approach of Ref.~\cite{Jac06} (see also
Refs.~\cite{Ric-book}).
We first write the transmission probabilities
as
\begin{eqnarray}\label{semicl-tr}
T_{ji}^{\beta \alpha} &=& \frac{1}{2 \pi \hbar} \int_i \rmd y_0 \int_j \rmd y
\sum_{\gamma1,\gamma2} A_{\gamma1} A_{\gamma2}^* \exp[i \delta S/\hbar]
\, .\qquad
\end{eqnarray} 
This expression sums over all  classical 
trajectories $\gamma1$ and $\gamma2$ entering the cavity 
at $y_0$ on a cross-section
of lead $i$ and exiting at $y$ on a cross-section of lead $j$, while
converting an $\alpha$ quasiparticle into a $\beta$ quasiparticle. 
The phase 
$\delta S = S_{\gamma1}-S_{\gamma2}$
gives the difference in action phase accumulated along $\gamma1$ and $\gamma2$.
In presence of tunnel barriers, the stability 
amplitude $A_\gamma$ is given by~\cite{Whi07,Cou92}
\begin{equation}\label{eq:stability}
A_\gamma = B_\gamma \, t_i t_j \,
\prod_k [r_k]^{l_{\gamma}(k)},
\end{equation}
where $l_\gamma(k)$ gives the number of times that $\gamma$ is reflected
back into the system from the tunnel barrier $k=L,R$, the transmission
and reflection amplitudes at the normal leads satisfy $|t_i|^2 = (1-|r_i|^2)=
\Gamma_{\rm n,s}$ (for $i=L, \, R, \, S_1,$ or $S_2$), and
$B_\gamma^2 = (d p_{y_0}/d y)_\gamma$ measures the rate of change of the 
initial momentum $p_{y_0}$ as the exit position $y$ of $\gamma$ is changed,
for a fixed sequence of transmissions and reflections at the tunnel
barriers.

We use Eqs.~(\ref{semicl-tr}) and (\ref{eq:stability})
to evaluate the contributions to the total
conductance, Eq.~(\ref{eq:sym_lambert}), arising from trajectories 
touching both superconductors such as 
those sketched in Fig.~\ref{fig:device}. These are the only trajectories that
are $\phi$-dependent, they are subdivided
into class I trajectories, contributing to 
$\langle T_{LL}^{he} \rangle $ (blue trajectory on Fig.\ref{fig:device}), 
and class II trajectories, contributing to $\langle T_{RL}^{ee} \rangle $
(red trajectory on Fig.\ref{fig:device}).
From now on, we focus our discussion on 
class I trajectories. The calculation of class II contributions proceeds
along the same lines and will be presented elsewhere~\cite{Goo08}.
Class I trajectories are made of the following sequence
\begin{equation}\label{eq:classI}
\gamma_I^{(p)} = \gamma^{(e)}_{s1}+\gamma^{(h)}_{s1}
+p \times \left[\gamma^{(h)}_{s3}+\gamma^{(e)}_{s3}+\gamma^{(e)}_{s1}+\gamma^{(h)}_{s1}
\right]  ,
\end{equation}
where $s1$ and $s3$ can be interchanged, and $p=0,1,2,\ldots$
They undergo $2p+1$ Andreev reflections, $2p$ reflections at 
tunnel barriers~\cite{caveat}, and accumulate an action phase
\begin{eqnarray}\label{eq:action_classI}
S_{\gamma,I} &=& p (-\pi -\phi + \epsilon \, t_{\ell \rm,I}) + 2 \epsilon \, t_{\gamma_{s1}}  
 - (\pi/2+\phi/2).
\end{eqnarray}
One should substitute $\phi\rightarrow -\phi$ when interchanging
segments $s1$ and $s3$, but the relative sign 
between $\pi$ and $\phi$ does not affect the final result.
Here, $t_{\ell \rm,I}$ gives the duration of the Andreev loop [the 
sequence between bracket 
in Eq.~(\ref{eq:classI})], $t_{\gamma_{s1}}$ the duration of the 
segment $\gamma_{s_1}$.
We see that at $\epsilon=0$ and 
$\phi=\pi$, the phase difference accumulated by any two members
(with different $p$ and $p'$) of a given family vanishes, so that
all pairs of trajectories within a given family 
resonate. There is however no resonance between
members of different families. 

In normal chaotic billiards, the stability $B_\gamma^2$ of
periodic orbits decreases exponentially with the number of times the
orbit is traveled~\cite{Haake-book}. The situation is fundamentally
different in presence of superconductivity, where Andreev reflections
refocus the dynamics. The stability of a trajectory is then 
given by the product of the stabilities
along the primitive segments ($\gamma_{s1}$ and $\gamma_{s3}$ for class I,
$\gamma_{s1}$, $\gamma_n$ and $\gamma_{s2}$ for class II)
that the trajectories
are made of, independent of $p$~\cite{caveat2}. This is true 
as long as half the duration of the Andreev loop is shorter than the Ehrenfest 
time $\tau_{\rm E}$, i.e. the time beyond which an initially narrow
wavepacket can no longer fit inside
a superconducting lead~\cite{Lod98,Jac03}. 
For a quantum dot of linear size $L_{\rm c}$ and Lyapunov exponent $\lambda$
(in absence of superconductivity), one has
$\tau_{\rm E} = \lambda^{-1} \ln [N_{\rm s}^2/k_{\rm F} L_{\rm c}]$, which
determines the relative measure of trajectories
contributing to macroscopic resonant tunneling,
together with the average time $\tau_{\rm D}$ between
two consecutive Andreev reflections.

We are now ready to evaluate the dominant contributions to conductance
close to resonance at
$\epsilon=0$ arising from class I trajectories. 
We start from Eq.~(\ref{semicl-tr}), and,
following the above considerations, we substitute
\begin{eqnarray}\label{semicl-tr-spa}
&&\sum_{\gamma1,\gamma2} A_{\gamma1} A_{\gamma2}^* [ \dots ]_{\gamma1,\gamma2} 
\longrightarrow  \\
&&\Gamma_{\rm n}^2
\sum_{\gamma = {\rm prim}} B_{\gamma}^2 \sum_{p,p' =0}^{\infty} 
(1-\Gamma_{\rm n})^{a(p+p')} \, \Gamma_{\rm s}^{p+p'+c} [ \dots ]_{\gamma,p,p'} 
\, .\nonumber
\end{eqnarray}
To obtain (\ref{semicl-tr-spa}), we paired
trajectories by class, noting that for a given class, all 
trajectories have the same stability
but differ only by the number of Andreev reflections at the superconductors
and normal reflections at the normal leads, and 
by the different
action phases they accumulate along
their Andreev loop. 
The sum over classes is then represented by a sum over 
primitive
trajectories, and the exponents $a=1$ and $c=1$ 
for class I are determined by the number of Andreev and
normal reflections in Eq.~(\ref{eq:classI}).
Reflection phases at the tunnel barriers do not appear because all
trajectories are traveled as many times by an electron as by a hole.
The evaluation of $\sum B_{\gamma}^2$
proceeds along the lines of Ref.~\cite{Jac06}, and details will be presented elsewhere~\cite{Goo08}. 
The resonant part of the conductance from class I and II contributions 
finally reads
\begin{subequations}
\label{eq:semiclG}
\begin{eqnarray} \label{eq:TLLhe}
\langle T_{LL}^{he} \rangle_{\rm r}  &=& \frac{\pi \Gamma_{\rm n}^2 N_{\rm n}}{4} 
\left(\frac{N_{\rm s}}{2 \Gamma_{\rm n} N_{\rm n} + 2 \Gamma_{\rm s} N_{\rm s}}
\right)^2 \\
& \times &\Big(1-(1+\tau_{\rm E}/\tau_{\rm D}) \exp[-\tau_{\rm E}/\tau_{\rm D}]\Big) \nonumber \\
& \times &\frac{\Gamma_{\rm s} }
{1-2 \, \Gamma_{\rm s} \, (1-\Gamma_{\rm n}) \cos[\pi-\phi]+\Gamma_{\rm s}^2 \, (1-\Gamma_{\rm n})^2} \, . \nonumber \\
\label{eq:TRLee}
\langle T_{RL}^{ee}\rangle_{\rm r} 
&=& \frac{\pi^2 \Gamma_{\rm n}^2 \, N_{\rm n}^2}{8 
N_{\rm s}} \left(\frac{N_{\rm s}}{2 \Gamma_{\rm n} N_{\rm n} + 2 \Gamma_{\rm s} N_{\rm s}}\right)^3 \\
& \times &
\Big(1-(1+\tau_{\rm E}/\tau_{\rm D}+\tau_{\rm E}^2/2\tau_{\rm D}^2) 
\exp[-\tau_{\rm E}/\tau_{\rm D}]\Big) \nonumber \\
&\times& 
\frac{1\, + \, \Gamma_{\rm s}^2 \, 
(1-\Gamma_{\rm n})^2}{1-2 \, \Gamma_{\rm s} \, (1-\Gamma_{\rm n})^2 \cos[\pi-\phi]+\Gamma_{\rm s}^2 \, (1-\Gamma_{\rm n})^4} . \nonumber
\end{eqnarray}
\end{subequations}
The sum of
Eqs.~(\ref{eq:TLLhe}) and (\ref{eq:TRLee}) gives
the dominant semiclassical contribution to the conductance. 
It exhibits the functional
dependence of resonant tunneling [compare to Eq.~(\ref{eq:res_tunnel})], where the resonance is however
always at the Fermi level, and is achieved by setting the phase
difference between the two superconductors at $\phi=\pi$. This resonance
condition is the same for all 
trajectories. This is
why the resonance is macroscopic, $\propto N_{\rm n}$, and not of order
one, as is the case for standard resonant tunneling in 
chaotic systems. In most 
instances, $\langle T_{LL}^{he} \rangle_{\rm r} \gg 
\langle T_{RL}^{ee}\rangle_{\rm r} $. Then the
resonance height at large $\tau_{\rm E}/\tau_{\rm D}$, small 
$\Gamma_{\rm n}$ and $\Gamma_{\rm s}=1$ 
is given by $G(\pi) \simeq \pi N_n/16$. Simultaneously,
the sharpness of the
resonance peak, measured by its width at half height,
is proportional to $\Gamma_{\rm n}$. We also note that the effect
disappears if the superconductors are poorly connected to the normal cavity,
$\Gamma_{\rm s} \rightarrow 0$, as should be.

The conductance is the sum of the semiclassical contributions,
Eqs.~(\ref{eq:semiclG}), and of quantum universal contributions.
We calculated the latter using
Nazarov's circuit theory~\cite{Naz94} and obtained
$G_{\rm nct}(\pi)/G_0 
= \Gamma_{\rm n} N_{\rm n}/2$~\cite{Goo08}.
In the tunneling regime, this is smaller than the
semiclassical contribution
by a factor $\propto \Gamma_{\rm n} \ll 1$, i.e. semiclassical contributions
enhance the conductance by a factor $\Gamma_{\rm n}^{-1} \gg 1$.

We briefly confirm our predictions numerically. 
We extend the open kicked rotator
of Refs.~\cite{Jac03,krot} to take into account both transport between
two normal leads and Andreev reflection at two superconducting terminals.
We construct a four-terminal scattering matrix from the Floquet operator of 
the kicked rotator as in Refs.~\cite{Jac03,krot}, and
use the method of Ref.~\cite{Bee95} to evaluate 
the exact expression for the conductance~\cite{Lam93},
\begin{equation}
\langle G \rangle /G_0 = T_{RL}^{ee}+T_{RL}^{he}+
2 \frac{T_{LL}^{he} T_{RR}^{he}
-T_{LR}^{he} T_{RL}^{he}}{T_{LL}^{he}+T_{RR}^{he}+T_{LR}^{he}+T_{RL}^{he}}.
\end{equation}
In our numerics, we restrict ourselves to
perfectly connected superconductors, $\Gamma_{\rm s} = 1$.
We average our data over ensembles of systems with 
fixed classical parameters -- such as 
the width of the leads, the strength of the tunnel barriers and
the size and chaoticity of the cavity -- but different chemical potentials or
lead positions. We focus our investigations on the semiclassical limit
$k_{\rm F} L_{\rm c} \rightarrow \infty$.

\begin{figure}[ht]
\begin{center}
\psfrag{G(phi)} {$G(\phi)$}
\psfrag{G(pi)} {$G(\pi)$}
\psfrag{G(pi)/G(0)} {$\mbox{ \hspace{-0.5cm} $G(\pi)/G(0)$}$}
\psfrag{phi/2pi} {$\phi/2 \pi$}
\psfrag{kfl} {$k_{\rm F} L_{\rm c}$}
\includegraphics[width=8.5cm]{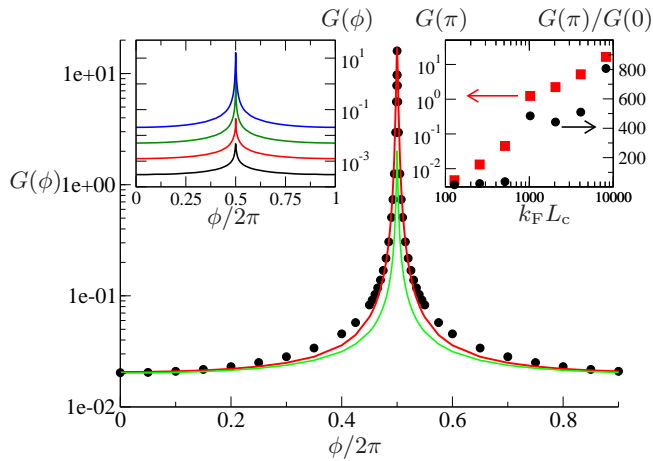}
\end{center}
\caption{
\label{fig:G_of_Phi} (Color online) 
Conductance through a chaotic Josephson junction
vs. the phase difference $\phi$ between the two superconductors. 
Circles are numerical results obtained from the kicked Josephson
rotator with $\Gamma_{\rm n}=0.01$, 
$k_{\rm F} L_{\rm c}=8192$, $k_{\rm F} L_{\rm c}/N_{\rm n}=20$, 
$k_{\rm F} L_{\rm c}/N_{\rm s}=10$ and
Lyapunov exponent $\lambda \simeq 1.3$. The red curve is the analytical
prediction obtained by summing the semiclassical
resonant contributions of Eqs.~(\ref{eq:semiclG}) with the
universal prediction. The green line gives the
universal prediction obtained from circuit theory.
Left inset: Numerical data 
for the same classical parameters
$k_{\rm F} L_{\rm c}/N_{\rm n}=20$, 
$k_{\rm F} L_{\rm c}/N_{\rm s}=10$ and $K=10$ as in the main plot, for
$k_{\rm F} L_{\rm c}=128$
(black curve), 512 (red curve), 2048 (green curve) and
8192 (blue curve). Note the change in peak-to-valley ratio.
Right inset: peak-to-valley ratio $G(\pi)/G(0)$ (black circles)
and peak conductance $G(\pi)$ (red squares) as a function of
$k_{\rm F} L_{\rm c}$, 
for the same classical configuration as in the main plot.
Data are averaged over 150--1000 sample realizations.\\[-8mm]}
\end{figure}

In Fig.~\ref{fig:G_of_Phi}, we show a resonance in the semiclassical
regime. We obtain very good agreement between
the numerical data (circles) and the analytical prediction (red solid line)
with $\tau_{\rm E}/\tau_{\rm D} \simeq 0.79$.
Without the semiclassical contribution, this agreement would
break down close to resonance, where universal contributions give a prediction
$G_{\rm nct}(\pi)=2$ (green line), too small by an order of magnitude.
The left inset illustrates the increase of the peak height and narrowness as
the semiclassical parameter $k_{\rm F} L_{\rm c}$ increases, all classical
parameters being kept constant. The four sets of data in this inset correspond
to a given classical configuration, with the electronic wavelength 
decreasing by factors of four from one curve to the next, starting from
the bottommost (black) curve. The conductance increases at each step because
the number of conduction channels scales linearly with $k_{\rm F} L_{\rm c}$.
In absence of semiclassical contributions, these
four curves would exhibit the same peak-to-valley ratio, but here they
obviously do not. This is 
quantified in the right inset to Fig.~\ref{fig:G_of_Phi},
where we show both the peak height and the peak-to-valley ratio
corresponding to the same configuration as in the main plot, 
while varying $k_{\rm F} L_{\rm c}$.

The connection can be made between the predicted and
observed enhancement of conductance at $\phi=\pi$ and
resonant tunneling through a macroscopic number of 
quasi-degenerate Andreev levels.
In $\pi$-biased closed chaotic Andreev billiards, Bohr-Sommerfeld
quantization predicts that all periodic orbits touching both superconductors
contribute to a peak in the density of states at the Fermi
energy with $\propto N_{\rm s}$ states. 
Once electrodes are connected to the billiard,
all those $\epsilon=0$ 
each level that significantly overlaps with the electrodes
contributes one perfect conductance 
channel to transport via resonant tunneling, which therefore
becomes macroscopic. We have numerically 
checked that the observed increase of conductance is accompanied
by the emergence of a large peak around $\epsilon=0$ in the
corresponding Andreev billiard. This and other results will be
presented elsewhere~\cite{Goo08}.

In summary, we investigated semiclassically the conductance through 
quantum chaotic Josephson junctions connected to two external normal
leads. We found an order-of-magnitude enhancement of the conductance
when the two superconductors have a phase difference of $\pi$. We identified
the mechanism behind this enhancement as resonant tunneling through
a macroscopic number of quasi-degenerate levels at the Fermi energy.

We thank C. Beenakker for drawing our attention to Ref.~\cite{Kad99},
and M. B\"uttiker for valuable comments on the manuscript.
M. Goorden was supported by the Swiss NSF.
P. Jacquod expresses his gratitude to M. B\"uttiker and the
Department of Theoretical Physics at the University of Geneva for their
hospitality.


\begin{thebibliography}{99}
\bibitem{caveat} The segments $\gamma_{n}$ and 
$\gamma_{si}$, $i=1,2,3$ in 
Fig.\ref{fig:device} might include many intermediate
bounces at a tunnel barrier. All possible number of reflections are 
effectively resummed; see Ref.~\cite{Whi07}.
\bibitem{caveat2} The contributions arising from the trajectories 
with $p=0$ are more stable than those with $p \ge 1$, because they do 
not travel on $\gamma_{s3}$. 
Their contribution is thus underestimated in our approach,
and Eqs.~(\ref{eq:semiclG}) underestimate the conductance by a 
subdominant correction $\propto \Gamma_{\rm n} N_{\rm n}$.
\end{thebibliography}
\end{document}